\documentclass[12pt, a4paper]{article}
\usepackage{amsmath}
\everymath{\displaystyle}
\usepackage{mathtools}
\usepackage{amsfonts}
\usepackage{amssymb}
\usepackage{amsthm}
\usepackage[utf8]{inputenc}
\usepackage[toc]{appendix}
\usepackage[hiresbb]{graphicx}
\usepackage{caption}
\usepackage{subcaption}
\usepackage{longtable}
\usepackage{booktabs}
\usepackage{listings}
\usepackage{here}
\usepackage{float}
\usepackage{ascmac}
\usepackage{tikz}
\usepackage{url}
\usepackage{lipsum}
\usepackage{authblk}
\usepackage{eqnarray}
\usepackage{siunitx}
\DeclareMathOperator\supp{supp}
\usepackage[sectionbib,round]{natbib}
\newtheorem{theorem}{Theorem}[section]

\newtheorem{definition}{Definition}[section]
\newtheorem{corollary}{Corollary}[section]

\DeclareMathOperator{\tr}{Tr}

\newcommand{\be}{\begin{equation}}
\newcommand{\ee}{\end{equation}}
\newcommand{\beq}{\begin{eqnarray*}}
\newcommand{\eeq}{\end{eqnarray*}}
\def\sym#1{\ifmmode^{#1}\else\(^{#1}\)\fi}

\title{\Large{\bf{Towards A Post-Quantum Cryptography in Blockchain I: Basic Review on Theoretical Cryptography and Quantum Information Theory
}}}
\author{\large{\bf{Tatsuru Kikuchi}}}
\affil{\small{\it{Faculty of Economics, The University of Tokyo,}}\\
{\it{7-3-1 Hongo, Bunkyo-ku, Tokyo 113-0033 Japan}}}
\date{\small{(\today)}}
\begin{document}
\maketitle
\begin{abstract}
Recently, the invention of quantum computers was so revolutionary that they bring transformative challenges in a variety of fields, especially for the traditional cryptographic blockchain, and it may become a real thread for most of the cryptocurrencies in the market. That is, it becomes inevitable to consider to implement a post-quantum cryptography, which is also referred to as quantum-resistant cryptography, for attaining quantum resistance in blockchains.
\end{abstract}
\newpage
\section{Introduction}
Recently, the invention of quantum computers was so revolutionary that they bring transformative challenges in a variety of fields, especially for the traditional cryptographic blockchain, and it may become a real thread for most of the cryptocurrencies in the market. That is, it becomes inevitable to consider to implement a post-quantum cryptography, which is also referred to as quantum-resistant cryptography, and post-quantum ledger for attaining quantum resistance in blockchains. 

One of the known quantum cryptography is the so-called, quantum key distribution (QKD). There exists a conceptually huge gap between the quantum key distribution and the public key cryptography (PKC). The security of key-encryption in QKD relies based on the mathematical foundation of quantum information theory, in contrast to traditional PKC that relies on the computational difficulty of the algebraic structure of elliptic curves over finite fields, which can not provide any mathematical proof as to the actual complexity of reversing the mathematical functions used. The key ingredient of QKD is based on the fundamental principle of quantum mechanics, in which the process of measuring a quantum system disturbs the system. If there is some third-party attacks to read the encrypted data, the quantum state in a Hilbert space, which is initially a superposition of several states in the Hilbert space, reduces to a single state that differs from the original state.

\section{Preliminary --- Cryptographic algorithms}
In this section, we describe the basic notions of cryptography. Cryptography is the foundations of secure communication including an exchange of transactions between two parties in the presence of adversaries.

\subsection{Encryption}
\begin{definition}[Encryption scheme]
Formally, an encryption scheme (or, cipher) is defined by a tuple ${\mathcal{E}} = ({\mathsf{Gen}}, {\mathsf{Enc}}, {\mathsf{Dec}})$ with a finite size message space ${\mathcal{M}}$. Each algorithms are defined as follows.
\begin{itemize}
\item
Key Generation Algorithm (${\mathsf{Gen}}$): it is a probabilistic algorithm that outputs a key $k$ chosen according to some distribution. The set of all possible key space generated by ${\mathsf{Gen}}$ is called a key space, denoted by ${\mathcal{K}}$. 
\item
Encryption Algorithm (${\mathsf{Enc}}$): it takes as input a key $k$ and a message $m$ and outputs a ciphertext $c$. We denote by ${\mathsf{Enc}}_{k}(m)$ the encryption of the plaintext $m$ using the key $k$. We denote by ${\mathcal{C}}$, the set of all possible ciphertext which is the output of ${\mathsf{Enc}}(k, m)$ for all possible choices of $k \in {\mathcal{K}}$ and $m \in {\mathcal{M}}$.
\item
Decryption Algorithm (${\mathsf{Dec}}$): it takes as input a key $k$ and a ciphertext $c$ and outputs a plaintext $m$. We denote the decryption of the ciphertext $c$ using the key $k$ by ${\mathsf{Dec}}(k, c)$. We assume the perfect correctness, meaning that for all $k \in {\mathcal{K}}$, $m \in {\mathcal{M}}$, and any ciphertext $c$, the output of ${\mathsf{Enc}}(k, m)$, it holds that $m = {\mathsf{Dec}}(k, c)$ with probability 1.
\end{itemize}
\end{definition}

\begin{definition}[Perfect secure]
Let ${\mathcal{E}} = ({\mathsf{Gen}}, {\mathsf{Enc}}, {\mathsf{Dec}})$ be an encryption scheme with a finite size message space ${\mathcal{M}}$, and consider a probabilistic experiment in which a random variable $k \in {\mathcal{K}}$ is uniformly distributed over ${\mathcal{K}}$. It is called perfect secure if for all messages $m_{0}, m_{1} \in {\mathcal{M}}$ and every ciphertext $c \in {\mathcal{C}}$, the following condition holds:
\be
{\mathbb{P}}\left({\mathsf{Enc}}(k, m_{0})=c \right) = {\mathbb{P}} \left({\mathsf{Enc}}(k, m_{1})=c \right) \;.
\ee
\end{definition}

\begin{theorem}
Let ${\mathcal{E}} = ({\mathsf{Gen}}, {\mathsf{Enc}}, {\mathsf{Dec}})$ be perfect secure encryption scheme with a finite size message space ${\mathcal{M}}$ and a key space ${\mathcal{K}}$, then we have $|{\mathcal{K}}| \geq |{\mathcal{M}}|$
\end{theorem}

\begin{definition}[One-time Pad]
One-time Pad is defined as an encryption scheme with the following property. For a given positive integer $\ell \in {\mathbb{N}}$, and an encryption scheme ${\mathcal{E}} = ({\mathsf{Gen}}, {\mathsf{Enc}}, {\mathsf{Dec}})$ with a finite size message space ${\mathcal{M}}$. where the keys, messages, and ciphertexts are the bit strings with the same length $\ell$, that is,
\be
{\mathcal{K}} = {\mathcal{M}} = {\mathcal{C}} = \{ 0, 1\}^{\ell} \;.
\ee
\begin{itemize}
\item
Key Generation Algorithm: it chooses a key from ${\mathcal{K}} = \{ 0, 1\}^{\ell}$ according to the uniform distribution.
\item
Encryption Algorithm: for a given key $k \in  \{ 0, 1\}^{\ell}$ and a message $m \in  \{ 0, 1\}^{\ell}$, it outputs the ciphertext $c = k \oplus m$.
\item
Decryption Algorithm: for a given key $k \in  \{ 0, 1\}^{\ell}$ and a ciphertext $c \in  \{ 0, 1\}^{\ell}$, it outputs the message $m = k \oplus c$.
\end{itemize}
\end{definition}

\begin{theorem}
One-time Pad is a perfect secure encryption scheme.
\end{theorem}

\begin{theorem}[Shannon's theorem]
Let ${\mathcal{E}} = ({\mathsf{Gen}}, {\mathsf{Enc}}, {\mathsf{Dec}})$ be an encryption scheme with a finite size message space ${\mathcal{M}}$, for which $|{\mathcal{K}}| = |{\mathcal{M}} | = | {\mathcal{C}} |$. The encryption scheme is perfect secure if and only if the following two properties hold,
\begin{itemize}
\item
Every key $k \in {\mathcal{K}}$ is chosen with equal probability ${\mathbb{P}}\left({\mathsf{Enc}}(k, m)=c \right) = 1/|{\mathcal{M}}|$ for every $m \in {\mathcal{M}}$ and $c \in {\mathcal{C}}$.
\item
For every $m \in {\mathcal{M}}$ and $c \in {\mathcal{C}}$, there exists a unique key $k \in {\mathcal{K}}$ such that ${\mathsf{Enc}}(k, m)=c$.
\end{itemize}
\end{theorem}

\begin{definition}[Attack Game 1  --- semantic security]
Let ${\mathcal{E}} = ({\mathsf{Gen}}, {\mathsf{Enc}}, {\mathsf{Dec}})$ be an encryption scheme defined over $({\mathcal{K}}, {\mathcal{M}}, {\mathcal{C}})$, and for a given adversary ${\mathcal{A}}$, we define two experiments described below. For $b=0, 1$, we consider the attack game as follows:
\begin{itemize}
\item
The adversary computes $m_{0}, m_{1} \in {\mathcal{M}}$, of the same length, and sends them to the challenger.
\item
The challenger computes $k \in {\mathcal{K}}$, $c = {\mathsf{Enc}}(k, m_{b})$, and sends $c$ to the adversary.
\item
The adversary outputs a bit $b \in \{0, 1 \}$.
\end{itemize}
For $b=0,1$, let $W_{b}$ be the event that the adversary ${\mathcal{A}}$ outputs $1$ in experiment $b$. We define the semantic security advantage of the adversary ${\mathcal{A}}$ with respect to ${\mathcal{E}} = ({\mathsf{Gen}}, {\mathsf{Enc}}, {\mathsf{Dec}})$ as
\be
{\mathsf{SSadv}}({\mathcal{A}},\, {\mathcal{E}}) = \big| {\mathbb{P}}\left(W_{0} \right) - {\mathbb{P}}\left(W_{1} \right) \big| \;. 
\ee
\end{definition}

\begin{definition}[Semantic security]
The cipher ${\mathcal{E}} = ({\mathsf{Gen}}, {\mathsf{Enc}}, {\mathsf{Dec}})$ is semantic secure if for all efficient adversaries ${\mathcal{A}}$, the value ${\mathsf{SSadv}}({\mathcal{A}},\, {\mathcal{E}})$ is negligible. 
\end{definition}

\begin{definition}[Attack Game 2 --- message recovery]
For a given cipher ${\mathcal{E}} = ({\mathsf{Gen}}, {\mathsf{Enc}}, {\mathsf{Dec}})$ be an encryption scheme defined over $({\mathcal{K}}, {\mathcal{M}}, {\mathcal{C}})$, and for a given adversary ${\mathcal{A}}$, the attack game proceeds as follows: 
\begin{itemize}
\item
The challenger computes $m \in {\mathcal{M}}$, $k \in {\mathcal{K}}$, $c = {\mathsf{Enc}}(k, m)$, and sends $c$ to the adversary.
\item
The adversary outputs a message $m^{\prime} \in {\mathcal{M}}$.
\item
Let $W$ be the event with $m^{\prime} = m$.
\end{itemize}
In this case, we say that the adversary ${\mathcal{A}}$ wins the game, and we define the message recovery advantage of the adversary ${\mathcal{A}}$ with respect to ${\mathcal{E}} = ({\mathsf{Gen}}, {\mathsf{Enc}}, {\mathsf{Dec}})$ as
\be
{\mathsf{MRadv}}({\mathcal{A}},\, {\mathcal{E}}) = \big| {\mathbb{P}}\left(W \right) - 1/|{\mathcal{M}}| \big| \;. 
\ee
\end{definition}

\begin{definition}[Message recovery security]
The cipher ${\mathcal{E}} = ({\mathsf{Gen}}, {\mathsf{Enc}}, {\mathsf{Dec}})$ is secure against message recovery if for all efficient adversaries ${\mathcal{A}}$, the value ${\mathsf{MRadv}}({\mathcal{A}},\, {\mathcal{E}})$ is negligible. 
\end{definition}

\begin{theorem}
Let ${\mathcal{E}} = ({\mathsf{Gen}}, {\mathsf{Enc}}, {\mathsf{Dec}})$ be a cipher defined over $({\mathcal{K}}, {\mathcal{M}}, {\mathcal{C}})$. If ${\mathcal{E}}$ is semantic secure then ${\mathcal{E}}$ is secure against message recovery.
\end{theorem}

\begin{definition}[Attack Game 3 --- parity prediction]
For a given cipher ${\mathcal{E}} = ({\mathsf{Gen}}, {\mathsf{Enc}}, {\mathsf{Dec}})$ be a cipher defined over $({\mathcal{K}}, {\mathcal{M}}, {\mathcal{C}})$. For $m \in {\mathcal{M}}$, we define parity by ${\mathrm{parity}}(m) =1$ if the number of $1$'s in $m$ is odd, otherwise ${\mathrm{parity}}(m) =0$. For a given adversary ${\mathcal{A}}$, the attack game proceeds as follows: 
\begin{itemize}
\item
The challenger computes $m \in {\mathcal{M}}$, $k \in {\mathcal{K}}$, $c = {\mathsf{Enc}}(k, m)$, and sends $c$ to the adversary.
\item
The adversary outputs $b^{\prime} \in \{0, \,1 \}$.
\item
Let $W$ be the event with $b^{\prime} = {\mathrm{parity}}(m)$.
\end{itemize}
In this case, we say that the adversary ${\mathcal{A}}$ wins the game, and we define the message recovery advantage of the adversary ${\mathcal{A}}$ with respect to ${\mathcal{E}} = ({\mathsf{Gen}}, {\mathsf{Enc}}, {\mathsf{Dec}})$ as
\be
{\mathsf{Parityadv}}({\mathcal{A}},\, {\mathcal{E}}) = \big| {\mathbb{P}}\left(W \right) - 1/2 \big| \;. 
\ee
\end{definition}

\begin{definition}[Parity prediction security]
A cipher ${\mathcal{E}} = ({\mathsf{Gen}}, {\mathsf{Enc}}, {\mathsf{Dec}})$ is secure against parity prediction if for all efficient adversaries ${\mathcal{A}}$, the value ${\mathsf{Parityadv}}({\mathcal{A}},\, {\mathcal{E}})$ is negligible. 
\end{definition}

\begin{theorem}
Let ${\mathcal{E}} = ({\mathsf{Gen}}, {\mathsf{Enc}}, {\mathsf{Dec}})$ be a cipher defined over $({\mathcal{K}}, {\mathcal{M}}, {\mathcal{C}})$. If ${\mathcal{E}}$ is semantic secure then ${\mathcal{E}}$ is secure against parity prediction.
\end{theorem}

\begin{definition}[Negligible function]
A function $f: {\mathbb{N}} \to {\mathbb{R}}$ is called negligible if for all $c \in \mathbb{R} _ {++}$ there exists $n_{0} \in {\mathbb{N}}$ such that for all integers $n \in {\mathbb{Z}}$ with $n \geq n_{0}$, we have $|f(n) | < 1/n^{c}$.
\end{definition}

\begin{corollary}
A function $f: {\mathbb{N}} \to {\mathbb{R}}$ is negligible if and only if for all $n \in {\mathbb{Z}}$ and $c \in \mathbb{R} _ {++}$, we have
\be
\lim_{n \to \infty} f(n) \, n^{c} = 0 \;.
\ee
\end{corollary}

\begin{definition}[Super-poly function]
A function $f: {\mathbb{N}} \to {\mathbb{R}}$ is called super-poly if $1/f$ is negligible.
\end{definition}

\begin{definition}[Poly-bounded function]
A function $f: {\mathbb{N}} \to {\mathbb{R}}$ is called poly-bounded if there exists $c, d \in \mathbb{R} _ {++}$ such that for all integers $n \in {\mathbb{Z}}$, we have $|f(n) | \leq n^{c} + d$.
\end{definition}

\begin{definition}[Efficient algorithm]
Let ${\mathcal{A}}$ be an efficient (probabilistic) algorithm that takes as input a security parameter $\lambda \in {\mathbb{N}}$ as well as other parameters encoded as a bit string $x \in \{ 0, 1\}^{\leq p(\lambda)}$ for some fixed polynomial function $p$. We call ${\mathcal{A}}$ an efficient (probabilistic) algorithm if there exists a poly-bounded function $\tau$ and a negligible function $\epsilon$ such that for all $\lambda \in {\mathbb{N}}$ and all $x \in \{ 0, 1\}^{\leq p(\lambda)}$, the probability that the running time of ${\mathcal{A}}$ on input $(\lambda, x)$ exceeds $\tau(\lambda)$ is at most $\epsilon(\lambda)$.
\end{definition}

\begin{definition}[System parameterization]
A system parameterization is an efficient probabilistic algorithm $P$ that given a security parameter $\lambda \in {\mathbb{N}}$ as input, outputs a bit string $\Lambda$, which is called a system parameter, whose length is bounded by a polynomial function in $\lambda$. We also define something related terminologies.
\begin{itemize}
\item
A collection ${\mathbb{S}} = \{ S_{\lambda \,,\Lambda} \}_{\lambda \,, \Lambda}$ of a finite sets of bits of strings, where $\lambda \in {\mathbb{N}}$ and $\Lambda \in \supp (P(\lambda))$, is called as a family of spaces with system parameterization $P$, provided the lengths of all the strings in each of the sets $S_{\lambda \,,\Lambda}$ are bounded by some polynomial function $p$ in $\lambda$. 
\item
We call that ${\mathbb{S}} = \{ S_{\lambda \,,\Lambda} \}_{\lambda \,, \Lambda}$ is efficiently recognizable if there is an efficient deterministic algorithm that on input $\lambda \in {\mathbb{N}}$, $\Lambda \in \supp (P(\lambda))$, and $s \in \{ 0, 1\}^{\leq p(\lambda)}$, determines if $\lambda \in S_{\lambda \,,\Lambda}$.
\item
We call that ${\mathbb{S}} = \{ S_{\lambda \,,\Lambda} \}_{\lambda \,, \Lambda}$ is efficiently sampleable if there is an efficient probabilistic algorithm that on input $\lambda \in {\mathbb{N}}$ and $\Lambda \in \supp (P(\lambda))$, outputs an element uniformly distributed over $S_{\lambda \,,\Lambda}$.
\item
We call that ${\mathbb{S}} = \{ S_{\lambda \,,\Lambda} \}_{\lambda \,, \Lambda}$ has an efficient length function if there is an efficient deterministic algorithm that on input $\lambda \in {\mathbb{N}}$, $\Lambda \in \supp (P(\lambda))$, and $s \in S_{\lambda \,,\Lambda}$, outputs a non-negative integer, called the length of $s$.
\end{itemize}
\end{definition}

\begin{definition}[Computational cipher]
A computational cipher consists of a pair of algorithms $({\mathsf{Enc}}, {\mathsf{Dec}})$, along with three families of spaces with system parameterization $P$. 
\beq
{\mathbb{K}} &=& \{ K_{\lambda \,,\Lambda} \}_{\lambda \,, \Lambda}\;, \\
{\mathbb{M}} &=& \{ M_{\lambda \,,\Lambda} \}_{\lambda \,, \Lambda}\;, \\
{\mathbb{C}} &=& \{ C_{\lambda \,,\Lambda} \}_{\lambda \,, \Lambda}\;.
\eeq
\begin{itemize}
\item
${\mathbb{K}}$, ${\mathbb{M}}$, and ${\mathbb{C}}$ are efficiently recognizable. 
\item
${\mathbb{K}}$ is efficiently sampleable.
\item
${\mathbb{M}}$ has an efficient length function.
\item
${\mathsf{Enc}}$ is an efficient probabilistic algorithm that on inputs $\lambda \,,\Lambda \,, k\,, m$, where $\lambda \in {\mathbb{N}}$, $\Lambda \in \supp (P(\lambda))$, $k \in K_{\lambda \,,\Lambda}$, and $m \in M_{\lambda \,,\Lambda}$, outputs an element of $C_{\lambda \,,\Lambda}$.
\item
${\mathsf{Dec}}$ is an efficient deterministic algorithm that on inputs $\lambda \,,\Lambda \,, k\,, c$, where $\lambda \in {\mathbb{N}}$, $\Lambda \in \supp (P(\lambda))$, $k \in K_{\lambda \,,\Lambda}$, and $c \in C_{\lambda \,,\Lambda}$, outputs an element of $M_{\lambda \,,\Lambda}$, or a special symbol which express to reject $\notin M_{\lambda \,,\Lambda}$.
\item
For all $\lambda \,,\Lambda \,, k\,, m\,,c$, where $\lambda \in {\mathbb{N}}$, $\Lambda \in \supp (P(\lambda))$, $k \in K_{\lambda \,,\Lambda}$, $m \in M_{\lambda \,,\Lambda}$, and $c \in \supp ({\mathsf{Enc}}(\lambda \,,\Lambda\,; k\,, m))$, we have $m = {\mathsf{Dec}}(\lambda \,,\Lambda\,; k\,, c)$. 
\end{itemize}
\end{definition}

\subsection{Stream cipher}
Recall that One-Time Pad is a Shanon's cipher ${\mathcal{E}} = ({\mathsf{Gen}}, {\mathsf{Enc}}, {\mathsf{Dec}})$ defined over $({\mathcal{K}}, \mathcal{M}, \mathcal{C})$, where $| {\mathcal{K}}| = | {\mathcal{M}}| = | {\mathcal{C}}| = \{0, 1 \}^{L}$. Here we consider the possible scheme with shorter length of encryption key. We consider the case where the encryption key is given by a $n$-bit string with $\ell < L$. Thus, the modified One-Time Pad is defined as follows: for $s \in \{0, 1 \}^{\ell}$ and $m, c \in \{0, 1 \}^{L}$, the encryption and decryption scheme are given by 
\be
{\mathsf{Enc}}(s, m) = G(s) \oplus m \;, ~ {\mathsf{Dec}}(s, c) = G(s) \oplus c \;,
\ee
where the function $G(s)$ is called as the pseudo-random generator (PRG), and this cipher is called as stream cipher. According to the Shannon's Theorem, this stream cipher by itself cannot achieve a perfect security; however if $G(s)$ satisfies some appropriate security property, then the stream cipher can be semantic secure. 

\begin{definition}[Pseudo-random generator (PRG)]
A pseudo-random generator (PRG) consists of an algorithm $G: \{0, 1 \}^{\ell} \to \{0, 1 \}^{L}$, along with two families of spaces with system parametrization $P$:
\be
{\mathbb{S}} = \{ S_{\lambda \,,\Lambda} \}_{\lambda \,, \Lambda} ~~{\mathrm{and}}~~{\mathbb{R}} = \{ R_{\lambda \,,\Lambda} \}_{\lambda \,, \Lambda}\;,
\ee
such that
\begin{itemize}
\item
${\mathbb{S}}$ and ${\mathbb{R}}$ are efficiently recognizable and sampleable.
\item
Algorithm $G$ is an efficient deterministic algorithm that on inputs $\lambda \,,\Lambda \,, s$, where $\lambda \in {\mathbb{N}}$, $\Lambda \in \supp (P(\lambda))$, and $s \in S_{\lambda \,,\Lambda}$, outputs an element $r \in R_{\lambda \,,\Lambda}$.
\end{itemize}
\end{definition}

\begin{definition}[Attack Game 4 --- PRG]
For a given PRG $G$, defined over $(S, R)$, and for a given adversary ${\mathcal{A}}$, the attack game proceeds as follows.
\begin{itemize}
\item
The challenger computes $r \in R$ as follows. \\
If $b=0$: $r = G(s)$ for $s \in S$, \\
If $b=1$: $r \in R$. \\
and sends the output $r$ to the adversary.
\item
The adversary outputs a bit $b^{\prime} \in \{0, 1 \}$.
\end{itemize}
Let $W_{b}$ be the event where the adversary ${\mathcal{A}}$ outputs a bit $1$ for a given bit $b \in \{0, \,1 \}$. We define ${\mathcal{A}}$'s advantage with respect to $G$ as
\be
{\mathsf{PRGadv}}({\mathcal{A}},\, G) = \big| {\mathbb{P}}\left(W_{0} \right) - {\mathbb{P}}\left(W_{1} \right) \big| \;. 
\ee
\end{definition}

\begin{definition}[PRG security]
A PRG $G$ is secure if the value ${\mathsf{PRGadv}}({\mathcal{A}},\, G)$ is negligible for all efficient adversaries ${\mathcal{A}}$.
\end{definition}

\begin{definition}[Stream cipher]
Let $G: \{0, 1 \}^{\ell} \to \{0, 1 \}^{L}$ be a PRG algorithm, the stream cipher ${\mathcal{E}} = ({\mathsf{Gen}}, {\mathsf{Enc}}, {\mathsf{Dec}})$ constructed from $G$ is defined over $(\{0, 1 \}^{\ell}, \{0, 1 \}^{\leq L}, \{0, 1 \}^{\leq L})$. For $s \in \{0, 1 \}^{\ell}$ and $m, c \in \{0, 1 \}^{\leq L}$, the encryption and decryption schemes are defined as follows. If $|m| = v$, 
\be
{\mathsf{Enc}}(s, m) = G(s)[0, \cdots, v-1] \oplus m \;,
\ee
and if $|c| = v$,
\be
{\mathsf{Dec}}(s, c) = G(s)[0, \cdots, v-1] \oplus c \;.
\ee
\end{definition}

\subsection{Block cipher}
\begin{definition}[Block cipher]
A block cipher is a deterministic cipher ${\mathcal{E}} = ({\mathsf{Gen}}, {\mathsf{Enc}}, {\mathsf{Dec}})$ whose message space and cipher space are given by the same set ${\mathcal{S}}$. We call $S$ as the data block space, and whose element $s \in S$ is referred as the data block.
\end{definition}

There is a famous product of block cipher, which is the so-called, AES (Advanced Encryption Standard). The AES keys (or, AES data block) has 128-bit strings, hence the size of data block space is given by $ | \mathcal{S}| = 2^{128}$. The key ingredient of block cipher is that we choose two steps for choosing random numbers: 1) random choice of keys, 2) random permutation of a chosen key. We denotes the set of all permutations on $\mathcal{S}$ as ${\mathsf{Perm{ [\mathcal{S} ]}}}$. The size of those set is given by $| {\mathsf{Perm{ [\mathcal{S} ]}}} | = | \mathcal{S}|!$. In the case of AES scheme with $ | \mathcal{S}| = 2^{128}$, the number of permutations is about 
\be
| {\mathsf{Perm{ [\mathcal{S} ]}}} | \cong 2^{2^{135}} \;,
\ee
while the number of permutations defined by 128-bit strings AES scheme is at most $2^{128}$. More precisely, block cipher is constructed to behave as pseudo-random permutations. Because there are $2^{\ell}!$ permutations on $\ell$-bit strings, it is said that a secure block cipher should be computational indistinguishable from a random permutation. 

\begin{definition}[Attack Game 5 --- block cipher]
For a given block cipher ${\mathcal{E}} = ({\mathsf{Gen}}, {\mathsf{Enc}}, {\mathsf{Dec}})$ defined over $({\mathcal{K}}, \mathcal{S}, \mathcal{S})$.
, and for a given adversary ${\mathcal{A}}$, we define two experiments described below. For $b=0, 1$, the attack game proceeds as follows:
\begin{itemize}
\item
The challenger choose a function $f$ as follows: \\
If $b=0$: $k \in \mathcal{K}$, $f(s) = {\mathsf{Enc}}(k, s)$ for $s \in \mathcal{S}$, else if $b=1$: $f \in \mathsf{Perm{ [\mathcal{S} ]}}$.
\item
The adversary send a sequence of queries to the challenger. \\
For a given $i \in {\mathbb{N}}$, $i$-th query is is a data block $x_{i} \in \mathcal{S}$. The challenger computes $y_{i} = f(x_{i}) \in \mathcal{S}$, and gives $y_{i}$ to the adversary. 
\item
The adversary outputs a bit $b^{\prime} \in \{0, \,1 \}$.
\end{itemize}
Let $W_{b}$ be the event where the adversary ${\mathcal{A}}$ outputs a bit $1$ for a given bit $b \in \{0, \,1 \}$. We define ${\mathcal{A}}$'s advantage with respect to the block cipher ${\mathcal{E}}$ as
\be
{\mathsf{BCadv}}({\mathcal{A}},\, {\mathcal{E}}) = \big| {\mathbb{P}}\left(W_{0} \right) - {\mathbb{P}}\left(W_{1} \right) \big| \;. 
\ee
Finally, we call that the adversary ${\mathcal{A}}$ is a q-query BC adversary if ${\mathcal{A}}$ issues at most q queries.
\end{definition}

\begin{definition}[Block cipher security]
A block cipher ${\mathcal{E}}$ is secure if for all efficient adversaries ${\mathcal{A}}$, the value ${\mathsf{BCadv}}({\mathcal{A}},\, {\mathcal{E}})$ is negligible.
\end{definition}

Let ${\mathcal{E}} = ({\mathsf{Gen}}, {\mathsf{Enc}}, {\mathsf{Dec}})$ be a block cipher defined over $({\mathcal{K}}, \mathcal{S}, \mathcal{S})$. If ${\mathcal{E}}$ is secure, it must be unpredictable, which means that every efficient adversary wins the following prediction game with negligible probability. In the prediction game, the challenger chooses a random key $k \in {\mathcal{K}}$, and the adversary send a sequence of queries $\{x_{1}, \cdots, x_{q} \}$ in response to the $i$-th query $x_{i} \in \mathcal{S}$ and the challenger responds with ${\mathsf{Enc}}(k, x_{i}) \in \mathcal{S}$. These queries are adaptive since each query may depend on the previous response. Finally, the adversary ${\mathcal{A}}$ outputs a pair of values $(x_{q+1}, \, y)$, where $x_{q+1} \notin \{x_{1}, \cdots, x_{q} \}$. The adversary ${\mathcal{A}}$ wins this prediction game if $y = {\mathsf{Enc}}(k, x_{q+1})$. 

Furthermore, if ${\mathcal{E}}$ is unpredictable, then it is secure against key-recovery, which means that every efficient adversary wins the following key-recovery game with negligible probability. In the key-recovery game, the adversary interacts with the challenger as the same with the prediction game, except that at the end, the adversary ${\mathcal{A}}$ outputs a candidate key $k^{\prime} \in {\mathcal{K}}$, and ${\mathcal{A}}$ wins the game if $k^{\prime} = k$. 

Combining those two implications, if  ${\mathcal{E}}$ is a secure block cipher, it must be secure against key-recovery. Moreover, if ${\mathcal{E}}$ is secure against key-recovery, it must be the case that $| {\mathcal{K}} |$ is large ({\it i.e.}, super-poly). We can see this as follows. An adversary can always win the key-recovery game with the probability $1/| {\mathcal{K}} |$ by simply choosing the key $k^{\prime} \in {\mathcal{K}}$ at random. If $| {\mathcal{K}} |$ is not super-poly, then this probability $1/| {\mathcal{K}} |$ is non-negligible. 

We can trade success probability for running-time proceeding a different attack, called an exhaustive search attack. In the exhaustive search attack, the adversary ${\mathcal{A}}$ makes a few, arbitrary queries $\{x_{1}, \cdots, x_{q} \}$ in the key-recovery game to obtain the responses $\{y_{1}, \cdots, y_{q} \}$. One may argue --- at least, assuming that $| {\mathcal{S}} | \leq | {\mathcal{K}} |$ and $| {\mathcal{K}} |$ is super-poly --- that for fairly small values of $q$ (in fact, $q=2$), with all but negligible probability, only one key satisfies
\be
y_{i} = {\mathsf{Enc}}(k, x_{i}) ~{\mathrm{for}} ~i = 1, \cdots, q \;.
\ee
If there is only one such key, the key that the adversary finds will be the one that is chosen by the challenger, and the adversary will win the game. Thus, the adversary wins the key-recovery game with all but negligible probability; however, the running time is linear in $| {\mathcal{K}}|$.

\begin{definition}[ECB cipher]
Let ${\mathcal{E}} = ({\mathsf{Gen}}, {\mathsf{Enc}}, {\mathsf{Dec}})$ be a block cipher defined over $({\mathcal{K}}, \mathcal{S}, \mathcal{S})$. For any poly-bounded $\ell \geq 1$, we can define a block cipher ${\mathcal{E}}^{\prime}$ defined over $({\mathcal{K}}, \mathcal{S}^{\leq \ell}, \mathcal{S}^{\leq \ell})$ as follows.
\begin{itemize}
\item
For $k \in {\mathcal{K}}$ and $m \in \mathcal{S}^{\leq \ell}$ with $v = | m|$, we define
\be
{\mathsf{Enc}}(k, m)^{\prime} = \left({\mathsf{Enc}}(k, m[0]), \cdots , {\mathsf{Enc}}(k, m[v-1])  \right) \;.
\ee
\item
For $k \in {\mathcal{K}}$ and $c \in \mathcal{S}^{\leq \ell}$ with $v = | c|$, we define
\be
{\mathsf{Dec}}(k, m)^{\prime} = \left({\mathsf{Dec}}(k, m[0]), \cdots , {\mathsf{Dec}}(k, m[v-1])  \right) \;.
\ee
We call this block cipher ${\mathcal{E}}^{\prime}$ as the $\ell$-wise electronic code book (ECB) cipher derived from the block cipher ${\mathcal{E}}$.
\end{itemize}
\end{definition}

\begin{theorem}
Let $\mathcal{E} = ({\mathsf{Gen}}, {\mathsf{Enc}}, {\mathsf{Dec}})$ be a block cipher defined over $(\mathcal{K}, \mathcal{S}, \mathcal{S})$. Let $\ell \in {\mathbb{N}}$ be any poly-bounded value, and ${\mathcal{E}}^{\prime}$ be the $\ell$-wise ECB cipher derived from ${\mathcal{E}}$, but with the message space restricted to all sequences of at most $\ell$ distinct data blocks. If ${\mathcal{E}}$ is a secure block cipher, then ${\mathcal{E}}^{\prime}$ is semantically secure. 
\end{theorem}

The block ciphers are the most primitive scheme in cryptography. In practice, constructing block ciphers use the same scheme, called the iterated block cipher. At first, we chooses a simple block cipher, $\mathcal{E} = ({\mathsf{Gen}}, {\mathsf{Enc}}, {\mathsf{Dec}})$ defined over $({\mathcal{K}}, \mathcal{S}, \mathcal{S})$, which is apparently insecure at this time. This block cipher is called as the round cipher. Second, we chooses a PRG $G$ that is used to generate $n$-sequence of pseudo-random keys $(k_{1}, \cdots, k_{n})$ from $k \in {\mathcal{K}}$. In this case, the PRG $G$ is called as the key expansion function. More precisely speaking, the second step in constructing the iterated block cipher is proceeded in iterative ways, that is,
\begin{itemize}
\item
Key expansion:
\be
(k_{1}, \cdots, k_{n}) = G(k) \;,~ {\mathrm{for}}~n \in {\mathbb{N}}\;,~ k \in  {\mathcal{K}} \;.
\ee
\item
Iteration: for each $i = 1, \cdots, n$, we apply the encryption function with the generated pseudo-random key $k_{i}$, and outputs
\be
y = {\mathsf{Enc}}(k_{n}, {\mathsf{Enc}}(k_{n-1}, \cdots, {\mathsf{Enc}}(k_{2}, {\mathsf{Enc}}(k_{1}, x))) \cdots) \;.
\ee
At each step, the operation of encryption function ${\mathsf{Enc}}$ is called the round, and the generated sequence of keys are called the round keys. The iteration of decryptions is almost the same except that the round keys are applied in reverse order.
\be
x = {\mathsf{Dec}}(k_{1}, {\mathsf{Dec}}(k_{2}, \cdots, {\mathsf{Dec}}(k_{n-1}, {\mathsf{Dec}}(k_{n}, y))) \cdots) \;.
\ee
\end{itemize}

One of the most famous algorithm of the round cipher is known as the Data Encryption Standard (DES). A strengthen version of the DES is called as the Triple-DES (3DES), which consists of three times of the rounds with each key space having $56$-bits key strings, hence the key length is $3 \times 56 = 168$. The Triple-DES has been approved through the year 2030 by the NIST as the U.S. standard in 1998. However, in 2002, it was superseded by the more secured AES algorithm, which has $128$-bits in each key space, and $10$ times of the rounds. 

\subsection{Universal hash function}
\begin{definition}[Hash function]
A hash function $H$ is a deterministic algorithm, whose inputs are a key $k$ and a message $m$, its output $t = H(k, m)$ is called as digest. The hash function is defined over the space $({\mathcal{K}}, \mathcal{M}, \mathcal{T})$, where the digest space $\mathcal{T}$ is the space in which the digest $t$ lies. In general, for two messages $m_{1}, m_{2} \in \mathcal{M}$ with $m_{1} \neq m_{2}$ and $k \in \mathcal{K}$, we say those two messages form collision for the hash function $H$ under the key $k$ if
\be 
H(k, m_{1}) = H(k, m_{2}) \;. 
\ee
\end{definition}

In general, since the size of the space $\mathcal{T}$ is much smaller than that of the space $\mathcal{M}$, it often happens to occur the collisions for the hash function $H$. However, it is necessary condition that the collision-less property of the hash function when using it in cryptography. In practice, it is enough to satisfy the weak collision-less property, in which the adversary must find a collision with no information on the keys at all. 

\begin{definition}[Attack Game 6 --- collision-less security]
For a hash function $H$ defined over the space $({\mathcal{K}}, \mathcal{M}, \mathcal{T})$, and consider the adversary ${\mathcal{A}}$, the attack game proceeds as follows. 
\begin{itemize}
\item
The challenger chooses a random key $k \in {\mathcal{K}}$, and keeps it to itself.
\item
The adversary ${\mathcal{A}}$ outputs two distinct messages $m_{1}, m_{2} \in \mathcal{M}$.
\end{itemize}
We say that the adversary ${\mathcal{A}}$ wins the game if two messages collide: 
\be
H(k, m_{1}) = H(k, m_{2}) \;. 
\ee
We define ${\mathcal{A}}$'s advantage with respect to $H$, denoted by ${\mathsf{UHFadv}}({\mathcal{A}},H)$ as the probability that ${\mathcal{A}}$ wins the game.
\end{definition}

\begin{definition}[Universal hash function]
Let $H$ be a hash function defined over the space $({\mathcal{K}}, \mathcal{M}, \mathcal{T})$.
\begin{itemize}
\item
We call that $H$ is $\epsilon$-bounded universal hash function (or $\epsilon$-UHF) if ${\mathsf{UHFadv}}({\mathcal{A}},H) \leq \epsilon$ for all the adversary ${\mathcal{A}}$. Equivalently, for every pair of distinct messages $m_{1}, m_{2} \in \mathcal{M}$, $H$ is $\epsilon$-UHF if we have
\be
{\mathbb{P}} \left[ H(k, m_{1}) = H(k, m_{2})   \right] \leq \epsilon \;,
\ee
where the probability is taken all over the random choice of $k \in {\mathcal{K}}$.
\item
We call that $H$ is a statistical UHF if it is an $\epsilon$-UHF with some negligible $\epsilon$.
\item
We call that $H$ is a computational UHF if ${\mathsf{UHFadv}}({\mathcal{A}},H)$ is negligible for all the efficient adversaries ${\mathcal{A}}$.
\end{itemize}
\end{definition}

\subsection{Public-key encryption}
\begin{definition}[Trapdoor function scheme]
A trapdoor function scheme is a tuple of efficient algorithms $\mathcal{E} = ({\mathsf{Gen}}, {\mathsf{Enc}}, {\mathsf{Dec}})$ along with two families of spaces with system parametrization $P$:
\be
{\mathbb{M}} = \{ M_{\lambda \,,\Lambda} \}_{\lambda \,, \Lambda} ~~{\mathrm{and}}~~{\mathbb{C}} = \{ C_{\lambda \,,\Lambda} \}_{\lambda \,, \Lambda}\;,
\ee
where $\lambda \in {\mathbb{N}}$ and $\Lambda \in \supp (P(\lambda))$. The following properties must be held.
\begin{itemize}
\item
${\mathbb{M}}$ is efficiently recognizable and sampleable.
\item
${\mathbb{C}}$ is efficiently recognizable.
\item
${\mathsf{Gen}}$ is an efficient probabilistic algorithm such that on inputs $\lambda \in {\mathbb{N}}$ and $\Lambda \in \supp (P(\lambda))$, it outputs a pair $(p_{k}, s_{k})$ where $p_{k}$ and $s_{k}$ are the bit strings those lengths must be poly-bounded in $\lambda$.
\item
${\mathsf{Enc}}$ is an efficient deterministic function so that on inputs $\lambda \in {\mathbb{N}}$, $\Lambda \in \supp (P(\lambda))$, $p_{k}$ with $(p_{k}, s_{k}) \in \supp ({\mathsf{Gen}}(\lambda, \Lambda))$, and $m \in M_{\lambda \,,\Lambda}$, it outputs an element of $C_{\lambda \,,\Lambda}$.
\item
${\mathsf{Dec}}$ is an efficient deterministic function so that on inputs $\lambda \in {\mathbb{N}}$, $\Lambda \in \supp (P(\lambda))$, $s_{k}$ with $(p_{k}, s_{k}) \in \supp ({\mathsf{Gen}}(\lambda, \Lambda))$, and $c \in C_{\lambda \,,\Lambda}$, it outputs an element of $M_{\lambda \,,\Lambda}$.
\item
For all $\lambda \in {\mathbb{N}}$, $\Lambda \in \supp (P(\lambda))$, $(p_{k}, s_{k}) \in \supp ({\mathsf{Gen}}(\lambda, \Lambda))$, and $m \in M_{\lambda \,,\Lambda}$, it satisfies
\be
{\mathsf{Dec}}(\lambda \,,\Lambda; s_{k}, {\mathsf{Enc}}(\lambda \,,\Lambda; p_{k}, m)) = m \;.
\ee
\end{itemize}
\end{definition}

One of the most famous example of the trapdoor function scheme is the so-called, RSA scheme. 
\begin{definition}[RSA trapdoor permutation scheme]
The basic algorithms for the RSA trapdoor permutation scheme $\mathcal{E} = ({\mathsf{Gen}}, {\mathsf{Enc}}, {\mathsf{Dec}})$ are explained as follows. It is parametrized by fixed values of $\ell >2$ (integer) and $e > 2$ (odd integer). Then, we proceed the following steps.
\begin{itemize}
\item
Key generation: At first, we use an efficient probabilistic algorithm ${\mathsf{RSAGen}}$ which outputs a pair of integers $(n, d) ={\mathsf{RSAGen}}(\ell, e)$, where $n = p \dot q$ with $p,~q$ be distinct primes, and $d = e^{-1}$ (mod $(p-1)(q-1)$). Then, a pair of keys $(p_{k}, s_{k})$ is given by $p_{k} = (n, e)$ and $s_{k} = (n, d)$.
\item
Encryption: For a given public key $p_{k} = (n, e)$ and $m \in {\mathbb{Z}}_{n}$, we define ${\mathsf{Enc}}(p_{k}, x) = x^{e} \in {\mathbb{Z}}_{n}$.
\item
Decryption: For a given secret key $s_{k} = (n, d)$ and $c \in {\mathbb{Z}}_{n}$, we define ${\mathsf{Dec}}(s_{k}, c) = c^{d} \in {\mathbb{Z}}_{n}$.
\end{itemize}
\end{definition}

\begin{definition}[Public key encryption]
A public key encryption scheme is a tuple $\mathcal{E} = ({\mathsf{Gen}}, {\mathsf{Enc}}, {\mathsf{Dec}})$ with the following properties, defined over the space $(\mathcal{M}, \mathcal{C})$.
\begin{itemize}
\item
${\mathsf{Gen}}$ is a probabilistic function which is defined by $(p_{k}, s_{k}) = {\mathsf{Gen}}(\kappa)~(\kappa \in \mathcal{K})$.
\item
For a given output of ${\mathsf{Gen}}$ and a message $m \in \mathcal{M}$, ${\mathsf{Enc}}$ is a probabilistic function which output the ciphertext $c ={\mathsf{Enc}}(p_{k}, m) \in \mathcal{C}$.
\item
${\mathsf{Dec}}$ is a deterministic function which is defined by $c = {\mathsf{Dec}}(s_{k}, m)$, where $c$ is a ciphertext and $m$ is either a message $m \in \mathcal{M}$ or a 'rejection' message that is distinct from all the messages.
\end{itemize}
\end{definition}

\begin{definition}[Attack Game 6 --- RSA security]
Let $\mathcal{E} = ({\mathsf{Gen}}, {\mathsf{Enc}}, {\mathsf{Dec}})$ be a RSA trapdoor permutation scheme with $\ell >2$ (integer) and $e > 2$ (odd integer). For a given adversary ${\mathcal{A}}$, the attack game proceeds as follows.
\begin{itemize}
\item
The challenger computes a pair of integers $(n, d) ={\mathsf{RSAGen}}(\ell, e)$, and outputs $m \in {\mathbb{Z}}_{n}$ and $c = m^{d} \in {\mathbb{Z}}_{n}$. Then, she sends the inputs $(n, c)$ to the adversary. 
\item
The adversary outputs $m^{\prime} \in {\mathbb{Z}}_{n}$.
\end{itemize}
We define ${\mathcal{A}}$'s advantage in breaking RSA trapdoor permutation scheme as the probability so that $m^{\prime} = m$, which is denoted by ${\mathsf{RSAadv}}({\mathcal{A}}, \ell, e)$.
\end{definition}

\begin{definition}[RSA assumption]
For given $\ell >2$ (integer) and $e > 2$ (odd integer), we call that the RSA assumption holds if for all ${\mathcal{A}}$, ${\mathsf{RSAadv}}({\mathcal{A}}, \ell, e)$ is negligible.
\end{definition}

\subsection{Key-exchange protocol}
The quotient group $({\mathbb{Z}}/n {\mathbb{Z}})^{\times}$ is defined by $({\mathbb{Z}}/n {\mathbb{Z}})^{\times} = \{0, \cdots, n-1\}$, whose order is given by the Euler's totient function $\varphi(n) = | ({\mathbb{Z}}/n {\mathbb{Z}})^{\times} |$. For prime $p$, $\varphi(p) = p-1$. The group $({\mathbb{Z}}/n {\mathbb{Z}})^{\times}$ becomes cyclic if and only if $n = 1, 2, 4, p^{k}$ or $n = 2 p^{k}$ where $p$ is odd prime and $k >0$. The quotient group becomes the cyclic group $C_{n}$, that is given by
\be
({\mathbb{Z}}/n {\mathbb{Z}})^{\times} \cong C_{\varphi(n)} \;, ~\text{where}~\varphi(p^{k}) = \varphi(2 p^{k}) = p^{k} - p^{k-1} \;.
\ee
By definition, the group becomes cyclic if and only if it has a generator $g$, that is, each element can be written by the powers of $g$:
\be
C_{\varphi(n)} = \{ g^{0} \, g^{1}, \cdots, g^{\varphi(n)-1} \} \;.
\ee
The Diffie-Hellman key-exchange protocol is briefly described as follows.
\begin{itemize}
\item
Alice computes $\alpha \in {\mathbb{Z}}_{q}$, $u = g^{\alpha} \in C_{p-1}$, and sends the output $u$ to Bob.
\item
Bob computes $\beta \in {\mathbb{Z}}_{q}$, $v = g^{\beta} \in C_{p-1}$, and sends the output $v$ to Alice.
\item
Alice computes $w = v^{\alpha}$ upon receiving $v$ from Bob.
\item
Bob computes $w = u^{\beta}$ upon receiving $u$ from Alice.
\end{itemize}
Then, the secret shared between them is given by
\be
w = v^{\alpha} = g^{\alpha \beta} = u^{\beta} \;.
\ee
It is said that the Diffie-Hellman key-exchange protocol is secure if and only if the following property holds:
\be
\text{For given}~g^{\alpha}, g^{\beta}  \in C_{p-1}\;,\text{where}~\alpha, \beta \in {\mathbb{Z}}_{q}, \text{it is hard to computes}~g^{\alpha \beta}  \in C_{p-1}\;.
\ee
This security property is called the computational Diffie-Hellman assumption.

\begin{definition}[Attack Game 7 --- decisional Diffie-Hellman security]
Let $C_{p-1}$ be a cyclic group with $p$ a prime, whose generated is written by $g \in C_{p-1}$. For a given adversary ${\mathcal{A}}$, the Diffie-Hellman attack game proceeds as follows. For given $b=0,1$:
\begin{itemize}
\item
The challenger computes 
$\alpha, \beta \in {\mathbb{Z}}_{q}$, $u = g^{\alpha}, v = g^{\beta} \in C_{p-1}$, and $g^{\alpha \beta}  \in C_{p-1}$ \;.
and send a pair of outputs $(u, v)$ to the adversary. For a given $b = 0, 1$,
\item
The adversary computes outputs some value of $w^{\prime} \in C_{p-1}$.
\item
Let $W_{b}$ be the event where the adversary ${\mathcal{A}}$ outputs a bit $1$ for a given bit $b \in \{0, \,1 \}$. We define $W_{1}$ be the event when the adversary ${\mathcal{A}}$ wins the game, {\it i.e.}, $w^{\prime} =w$. 
\be
{\mathsf{DDHadv}}({\mathcal{A}},\, C_{p-1}) = \big| {\mathbb{P}}\left(W_{0} \right) - {\mathbb{P}}\left(W_{1} \right) \big| \;. 
\ee
\end{itemize}
We define ${\mathcal{A}}$'s advantage in solving the decisional Diffie-Hellman problem with $C_{p-1}$. 
\end{definition}

\begin{definition}[Decisional Diffie-Hellman (DDH) assumption]
We say that decisional Diffie-Hellman (DDH) assumption holds for $C_{p-1}$ if for all efficient adversaries ${\mathcal{A}}$ the value of ${\mathsf{DDHadv}}({\mathcal{A}},\, C_{p-1})$ is negligible.
\end{definition}

\newpage
\section{Quantum Cryptography}
\subsection{Qubits, Entanglement, and Quantum Gates}
At the beginning, we summarize basic notions in describing for the quantum computers and quantum cryptography. In classical computers, the fundamental unit is a 'bit', which can be either 0 or 1. In quantum computers, the fundamental unit is a 'qubit', which is a superposition of 0 and 1. In general, a single qubit state is written as a state in the Hilbert space ${\cal{H}} = {\mathbb{C}}^{2}$
\be
\big| \psi \rangle = \alpha \, \big| 0 \rangle + \beta \, \big| 1 \rangle \;,
\ee 
where $|\alpha|^{2} + |\beta|^{2} = 1$ with the basis ${\mathcal{B}}_{z} = \left\{ \big| 0 \rangle, \, \big| 1 \rangle \right\}$. We can always make change of a basis to another one by using a unitary transformation. The following basis are also often used
\be
\big| \pm \rangle = \frac{\big| 0 \rangle \pm  \big| 1 \rangle}{\sqrt{2}} \;,
\ee
and we denote it as ${\mathcal{B}}_{x} = \left\{ \big| + \rangle, \, \big| - \rangle \right\}$.

We can construct a $N$ qubit system from single qubit systems by taking $N$ tensor products of the Hilbert spaces as ${\cal{H}}^{\otimes N} = \otimes_{i=1}^{N} {\cal{H}}_{i}$. 
For instance, 2 qubit state is spanned by 3 basis states when taking the orthogonal conditions into account. That is, taking a tensor product of two Hilbert spaces ${\cal{H}}_{1} \otimes {\cal{H}}_{2} =  {\mathbb{C}}^{2} \otimes {\mathbb{C}}^{2}$, the 2 qubit state can be expressed by
\be
\big| \psi_{1} \rangle \otimes \big| \psi_{2} \rangle = \big| \psi_{1} \psi_{2} \rangle \in {\cal{H}}_{1} \otimes {\cal{H}}_{2} \;,
\ee
where $\big| \psi_{1} \rangle  \in  {\cal{H}}_{1}$ and $\big| \psi_{2} \rangle  \in  {\cal{H}}_{2}$. We will often abbreviate $\big| \psi_{1} \rangle \otimes \big| \psi_{2} \rangle$ to $\big| \psi_{1} \psi_{2} \rangle $, etc.

An important property in quantum information theory is the so-called, 'entanglement', which refers to the quantum correlation between different qubits. In order to give more formal definition to that, we define a density operator $\rho$ associated to a given qubit $\big| \psi \rangle$ as follows.
\be
\rho = \big| \psi  \rangle \langle \psi   \big| \;.
\ee
The qubit is called a pure state if and only if the density operator satisfies the following condition,
\be
\rho^{2} = \rho \;.
\ee
Equivalently, a qubit state in the Hilbert space ${\cal{H}}_{1} \otimes {\cal{H}}_{2}$ is called a pure state if it can be expressed in the separable form: $\big| \psi_{1} \rangle \otimes \big| \psi_{2} \rangle$. On the other hand, a qubit state which cannot be written in such a separable way is called the entangled state. The standard basis for the two qubits consists of four orthonormal states which is called as the Bell states. The Bell states can be expressed as follows.
\beq
\big| \Psi_{1}^{\pm} \rangle &=& \frac{ \big| 01 \rangle \pm  \big| 10 \rangle}{\sqrt{2}} \;, \\
\big| \Psi_{2}^{\pm} \rangle &=& \frac{ \big| 00 \rangle \pm  \big| 11 \rangle}{\sqrt{2}} \;. 
\eeq
These states are maximally entangled. 

A unitary operator that acts on a small number of qubits (say, up to 3) is called 'quantum gate' (or, gate). That is a quantum analogous to the classical gates like 'AND', 'OR', and 'NOT'. 
If we take a set of basis in the Hilbert space ${\cal{H}}$, the quantum gates are defined by the outer products
\beq
\big| 0 \rangle \langle 0  \big| &=& 
\begin{pmatrix} 
1 & 0  \\ 
0 & 0 
\end{pmatrix}
\,,~~
\big| 0 \rangle \langle 1  \big| = 
\begin{pmatrix} 
0 & 1  \\ 
0 & 0 
\end{pmatrix} \;,
\\
\big| 1 \rangle \langle 0  \big| &=& 
\begin{pmatrix} 
0 & 0  \\ 
1 & 0 
\end{pmatrix}
\,,~~
\big| 1 \rangle \langle 1  \big| = 
\begin{pmatrix} 
0 & 0  \\ 
0 & 1 
\end{pmatrix} \;.
\eeq
The Pauli matrices are examples of $1$-qubit gates. 
\beq
{\mathsf{I}} &=& \big| 0 \rangle \langle 0  \big| + \big| 1 \rangle \langle 1  \big| \;,\\
{\mathsf{X}} &=& \big| 0 \rangle \langle 1  \big| + \big| 1 \rangle \langle 0  \big| \;,\\
{\mathsf{Y}} &=&  -i \big| 0 \rangle \langle 1  \big| + i  \big| 1 \rangle \langle 0  \big|  \;,\\
{\mathsf{Z}} &=& \big| 0 \rangle \langle 0  \big| -   \big| 1 \rangle \langle 1  \big| \;.
\eeq
Among those quantum gates, the bit-flip gate ${\mathsf{X}}$ ({\it a.k.a.}, NOT-gate) negates the computational basis, {\it i.e.}, it swaps $\big| 0 \rangle$ and $\big| 1 \rangle$. The phase-flip gate ${\mathsf{Z}}$ puts a minus sign in front of a qubit state $\big| 1 \rangle$. Another important $1$-qubit gate is the phase-shift gate, ${\mathsf{R}}_{\phi}$, which merely rotate the phase of the qubit state $\big| 1 \rangle$ by $\phi$, that is, ${\mathsf{R}}_{\phi} \big| 1 \rangle = e^{i \phi} \big| 1 \rangle$. Note that ${\mathsf{Z}}$ is a special case of ${\mathsf{R}}_{\phi}$ with $\phi=\pi$. Moreover, ${\mathsf{R}}_{\phi}$ gate with $\phi=\pi/4$ is often called as $T$-gate. In practice, the most relevant $1$-qubit gate is the so-called, Hadamard gate ${\mathsf{H}}$, which is given by
\be
{\mathsf{H}} = 
\frac{1}{\sqrt{2}} \,
\begin{pmatrix} 
1 & 1  \\ 
1 & -1 
\end{pmatrix}
\ee
The actions of the Hadamard gate ${\mathsf{H}}$ on the qubits exchange one basis to the other.
\beq
{\mathsf{H}} \, \big| 0 \rangle &=& \big| + \rangle \;, \quad {\mathsf{H}} \, \big| 1 \rangle = \big| - \rangle \;,\\
{\mathsf{H}} \, \big| + \rangle &=& \big| 0 \rangle \;,\quad{\mathsf{H}} \, \big| - \rangle = \big| 1 \rangle \;.
\eeq
The crucial property of the Hadamard gate is that if we act it to the qubit state with a superposition $\big| 0 \rangle + \big| 1 \rangle$, then we have
\be
{\mathsf{H}} \, \frac{1}{\sqrt{2}} \left( \big| 0 \rangle + \big| 1 \rangle    \right) =  \big| 0 \rangle \;.
\ee
The result of disappearance in the initial qubit state $\big| 1 \rangle$ is a result of the quantum interference. 

An example of a $2$-qubit gate is the {\it controlled-not} gate, CNOT. The CNOT gate negates the second bit of its input if the first qubit is 1, and does nothing if the first qubit is 0. 
\beq
{\mathsf{CNOT}} \, \big| 0 \rangle \otimes \big| b \rangle &=& \big| 0 \rangle \otimes \big| b \rangle \;,\\
{\mathsf{CNOT}} \, \big| 1 \rangle \otimes \big| b \rangle &=& \big| 0 \rangle \otimes \big| 1- b \rangle \;.
\eeq
where $b = 0, 1$. The first qubit is called as the control qubit and the second qubit is called as the target qubit, respectively. We can express the CNOT gate as a matrix form by
\be
{\mathsf{CNOT}} = 
\begin{pmatrix} 
1 & 0 & 0 & 0  \\ 
0 & 1 & 0 & 0  \\ 
0 & 0 & 0 & 1  \\
0 & 0 & 1 & 0  \\  
\end{pmatrix} \;.
\ee

\subsection{Quantum Teleportation}
Now, we consider the quantum transportation, in which Alice has a qubit $\big| \psi \rangle$ at the beginning, and she send the qubit to Bob. Let us consider the situation where an entangled pair of qubit is shared between Alice and Bob. Then, Alice performs a Bell measurement of the entangled pair of qubit and a target qubit $\big| \psi \rangle$, which is given by 
\be
\big| \psi \rangle = \alpha \big| 0 \rangle +  \beta \big| 1 \rangle \;,
\ee
and suppose that they share one of the Bell states, 
\be
\big| \Psi_{2}^{+} \rangle = \frac{ \big| 00 \rangle +  \big| 11 \rangle}{\sqrt{2}} \;.
\ee
Then, the total state is expressed by the tensor product between $\big| \psi \rangle$ and $\big| \Psi_{2}^{+} \rangle$ as follows.
\beq
\big| \psi \rangle \otimes \big| \Psi_{2}^{+} \rangle &=&
\frac{1}{2} \left(  \alpha \big| 000 \rangle + \beta \big| 100 \rangle + \alpha \big| 011 \rangle + \beta \big| 111 \rangle \right) \\
&=&
\frac{1}{2} \left\{  
\big| \Psi_{2}^{+} \rangle \otimes \left( \alpha \big| 0 \rangle +  \beta \big| 1 \rangle  \right)
+
\big| \Psi_{2}^{-} \rangle \otimes \left( \alpha \big| 0 \rangle -  \beta \big| 1 \rangle  \right) \right\} \\
&+&
\frac{1}{2} \left\{ 
\big| \Psi_{1}^{+} \rangle \otimes \left( \alpha \big| 0 \rangle +  \beta \big| 1 \rangle  \right)
+
\big| \Psi_{1}^{-} \rangle \otimes \left( \alpha \big| 0 \rangle -  \beta \big| 1 \rangle  \right) 
\right\} \;.
\eeq
When Alice makes the Bell measurement, then she will have one of the four Bell states, that can be encoded by two classical bits by 
\be
\Psi_{2}^{+}: 00\;, ~\Psi_{2}^{-}: 01\;, ~\Psi_{1}^{+}: 10\;, ~\Psi_{1}^{-}: 11\;.
\ee
Depending on the results of the Bell measurement, Bob performs one of four actions to his qubit and end up with the qubit $\big| \phi \rangle$. 
\begin{center}
\begin{tabular}{|c | c | c | c |}
\hline
Measurement & Classical bit & Bob's qubit & Quantum gate \\
\hline \hline
$\Psi_{2}^{+}$ & $00$ & $\alpha \big| 0 \rangle +  \beta \big| 1 \rangle$ & ${\mathsf{I}}$ \\
$\Psi_{2}^{-}$  & $01$ & $\alpha \big| 0 \rangle +  \beta \big| 1 \rangle$ & ${\mathsf{Z}}$ \\
$\Psi_{1}^{+}$ & $10$ & $\alpha \big| 0 \rangle +  \beta \big| 1 \rangle$ & ${\mathsf{X}}$ \\
$\Psi_{1}^{-}$  & $11$ & $\alpha \big| 0 \rangle +  \beta \big| 1 \rangle$ & ${\mathsf{Y}}$ \\
\hline
\end{tabular}
\end{center}
In every case, Bob can end up with the correct qubit $\big| \psi \rangle$ and the initial qubit $\big| \psi \rangle$ for Alice has collapsed after the Bell measurement. This process for sending a qubit from Alice to Bob is called as quantum transportation.

Let us consider a quantum gate $\rho$, we can construct a von Neumann entropy $S(\rho)$, which is defined by
\be
S(\rho) = - \tr \left( \rho \ln \rho  \right) \;.
\ee
The basic properties of von Neumann entropy are summarized below.
\begin{itemize}
\item $
S(\rho) \geq 0 \;,
$
\item $
S(\rho)  = 0 \;, ~\text{if and only if $\rho$ is a pure state} \;,
$
\item $
S(\rho) \leq N \;,~~\text{$N$ is the dimension of the Hilbert space: $N={\mathrm{dim}}({\cal{H}})$} \;.
$
\end{itemize}
In regard to the third property, the equality holds if and only if the qubit is maximally entangled. This means that a maximally entangled qubit is the state which maximize the von Neumann entropy. If we consider the tensor product Hilbert space ${\cal{H}} = {\cal{H}}_{1} \otimes {\cal{H}}_{2}$, and $\rho_{i}~(i = 1, 2)$ be the density operators on each Hilbert space. Then, we can define the relative (von Neumann) entropy by
\beq
S(\rho_{1} \| \rho_{2}) &=& \tr \left( \rho_{1} \ln \rho_{1} \right) - \tr \left( \rho_{1} \ln \rho_{2} \right) \\
&=& 
\tr \left[ \rho_{1} \left( \ln \rho_{1} - \ln \rho_{2} \right) \right] \;.
\eeq

\begin{theorem}[Holevo's theorem]
Let us consider an $N$ tensor products Hilbert space ${\cal{H}} = {\cal{H}}_{1} \otimes  \cdots \otimes {\cal{H}}_{N}$, and $\rho_{i} = \big| \psi_{i}  \rangle \langle \psi_{i} \big| $ where $\psi_{i} \in {\cal{H}}_{i}~(i = 1, \cdots , N)$. Suppose that any $\rho_{i}~(i = 1, \cdots , N)$ are the entangled states, and let $X$ be a random variable with $p_{i} = {\mathbb{P}}(X=i) ~(i = 1, \cdots , N)$, and define the total density operator $\rho$ as
\be
\rho = \sum_{i=1}^{N} p_{i} \, \rho_{i} \;,
\ee
and $Y$ be the Bell measurement on $\rho$. Then, it holds the following inequality
\be
S(\rho_{XY} \| \rho_{X} \otimes \rho_{Y}) \leq \chi \;, 
\ee
where $\chi = S(\rho) - \sum_{i=1}^{N} p_{i} \, S(\rho_{i})$, and $\chi$ is called the Holevo information.
\end{theorem}
This theorem states that there is an upper bound of the information that can be known on a quantum state when we perform a Bell measurement.

\subsection{Quantum Key Distribution}
The goal of Quantum Key Distribution (QKD) is to provide a way to share the secrete key between two parties, Alice and Bob. The ${\mathsf{BB84}}$ protocol is one of the first examples to realize the QKD. The basic procedures of ${\mathsf{BB84}}$ is described as follows. 
\begin{enumerate}
\item
Alice chooses a random bit and random basis using a random generator. 
\item
Alice encodes the chosen bit and sends the qubit to Bob by using a quantum channel. 
\item
Bob chooses a random basis using a random generator. 
\item
Bob measures the received qubit in the chosen basis and decodes the bit. 
\item
If Alice and Bob have the same basis, then they can share the same bit, and if not the case, they can only share the same bit withe a probability of 1/2, and they can share a different bit with the same probability of 1/2. 
\item
Alice and Bob repeat the steps 1 to 5 until a reasonable amount of bits have been exchanged. 
\item
After that they share the basis used for encoding and decoding for using the classical channel. 
\item
They discard every bit where the basis are not the same one, and keeps the others without revealing the value of bits. 
\item
To verify that nobody eavesdropped, they publicly share a reasonable amount of bits and verify that they agree. If they agree on all the bits, they discard the used bits and keep the secret key. If they disagree, they discard the secret key and start again.
\end{enumerate}
At the end of this procedure, Alice and Bob share the identical and secured secrete key. This ${\mathsf{BB84}}$ protocol has a loophole when the third part, Eve attacks to gain information on the exchanged bits. Before explaining the loophole of the ${\mathsf{BB84}}$ protocol, let us see the example of this procedure and how this protocol protects against passive eavesdropping. In the example shown below, we abbreviates the names of three characters by Alice, Bob, and Eve.

\begin{table}[h!]
\centering
\begin{tabular}{|c | c | c | c | c | c | c | c | c | c |}
\hline
n & A's bit & A's basis & A's qubit & E's basis & E's qubit & B's bit & B's basis & B's qubit & Key \\
\hline \hline
1 & 0 & ${\mathsf{Z}}$ & $\big| 0 \rangle$ & ${\mathsf{Z}}$ & $\big| 0 \rangle$ & 0 & ${\mathsf{Z}}$ & $\big| 0 \rangle$ & 0  \\
2 & 0 & ${\mathsf{Z}}$ & $\big| 0 \rangle$ & ${\mathsf{Z}}$ & $\big| 0 \rangle$ & 1 & ${\mathsf{X}}$ & $\big| - \rangle$ &   \\
3 & 1 & ${\mathsf{X}}$ & $\big| - \rangle$ & ${\mathsf{Z}}$ & $\big| 1 \rangle$ & 0 & ${\mathsf{X}}$ & $\big| + \rangle$ & 0  \\
4 & 0 & ${\mathsf{X}}$ & $\big| + \rangle$ & ${\mathsf{X}}$ & $\big| + \rangle$ & 1 & ${\mathsf{Z}}$ & $\big| 1 \rangle$ &   \\
5 & 1 & ${\mathsf{X}}$ & $\big| - \rangle$ & ${\mathsf{X}}$ & $\big| - \rangle$ & 1 & ${\mathsf{X}}$ & $\big| - \rangle$ & 1  \\
6 & 1 & ${\mathsf{Z}}$ & $\big| 1 \rangle$ & ${\mathsf{Z}}$ & $\big| 1 \rangle$ & 1 & ${\mathsf{Z}}$ & $\big| 1 \rangle$ & 1  \\
7 & 0 & ${\mathsf{Z}}$ & $\big| 0 \rangle$ & ${\mathsf{Z}}$ & $\big| 0 \rangle$ & 1 & ${\mathsf{X}}$ & $\big| - \rangle$ &   \\
8 & 0 & ${\mathsf{Z}}$ & $\big| 0 \rangle$ & ${\mathsf{X}}$ & $\big| + \rangle$ & 0 & ${\mathsf{Z}}$ & $\big| 0 \rangle$ & 0  \\
\hline
\end{tabular}
\caption{An example of ${\mathsf{BB84}}$ protocol}
\end{table}
If both Alice and Bob have the same basis, which correspond to $n = 1,3,5,6,8$, then they can share the same key, hence we only consider the case when $n = 1,3,5,6,8$. Let us denote ${\mathcal{B}}_{A}$, ${\mathcal{B}}_{A}$, and ${\mathcal{B}}_{A}$ the basis of Alice, Eve, and Bob, respectively. Lets consider the case when ${\mathcal{B}}_{A} = {\mathcal{B}}_{B} \neq {\mathcal{B}}_{E}$. Then, without loss of generality, we can take ${\mathcal{B}}_{A} = {\mathcal{B}}_{B} = {\mathsf{Z}}$ and ${\mathcal{B}}_{E} = {\mathsf{X}}$ and that Alice ant to encode her qubit $0$. In this case, the probability to find Eve is $1/2$. After discarding the bits, we have only two possibilities for the basis, 
\begin{enumerate}
\centering
\item
${\mathcal{B}}_{A} = {\mathcal{B}}_{B} = {\mathcal{B}}_{E} ; ~~p=\frac{1}{2}$
\item
${\mathcal{B}}_{A} = {\mathcal{B}}_{B} \neq {\mathcal{B}}_{E} ; ~~p=\frac{1}{2}$
\end{enumerate}
Then, the probability not to find her by revealing one bit becomes
\beq
P 
&=& {\mathbb{P}}({\mathcal{B}}_{A} = {\mathcal{B}}_{B} = {\mathcal{B}}_{E}) + 
{\mathbb{P}}({\mathcal{B}}_{A} = {\mathcal{B}}_{B} \neq {\mathcal{B}}_{E}) \times {\mathbb{P}}(\text{A's bit} = \text{B's bit})
\\
&=&
\frac{1}{2} + \frac{1}{2} \times \frac{1}{2} = \frac{3}{4} \;. 
\eeq
As the events are supposed to be independent each other, the probability not to find her by giving away $k$-bits is given by
\be
P_{k} = \left( \frac{3}{4}  \right)^{k} \;,
\ee
and then the probability not to find her by giving away $k$-bits becomes
\be
\hat{P}_{k} = 1-  \left( \frac{3}{4}  \right)^{k} \;,
\ee
The result shows that the more bits are giving away, the more chance to find Eve.

\section{Cryptographic Algorithms}
The primary cryptographic algorithms of blockchain basically consist of three types of cryptographic algorithms: cryptographic hash functions, symmetric cryptographic algorithms, and asymmetric cryptographic algorithms. The basic difference between symmetric cryptographic algorithms and asymmetric cryptographic algorithms is in the way to distribute the cryptographic keys. The use of Federal Information Processing Standard (FIPS) is approved by the security of commerce. The National Institute of Standards and Technology (NIST) has published a Special Publication (SP) which gives a standardization procedure as similar to the FIPS provided by the security of commerce.

\subsection{Cryptographic Hash Algorithms}
A cryptographic hash algorithm is a cryptographic primitive algorithm that basically map input data of any length to fixed length values. There are basically three types of security properties in the cryptographic hash algorithm (or hash function $H$). 
\begin{enumerate}
\item \textbf{One-way function}: \\
It is practically impossible to reverse the process from the hash value to the input value such that $H(x) = y$. 

\item \textbf{Weak-collision resistance}: \\
It is practically impossible to find another distinct value of $x^{\prime}$ for a given input value of $x$, such that the output of hash function match $H(x) = H(x^{\prime})~(x \neq x^{\prime})$.

\item \textbf{Strong-collision resistance}: \\
It is practically impossible to find two distinct values of $x$ and $x^{\prime}$ in such a way that the output of hash function match $H(x) = H(x^{\prime})$.
\end{enumerate}

\subsection{Symmetric Cryptographic Algorithms}
Symmetric cryptographic algorithm (or secret-key algorithm) uses a single, unique secrete-key in both encryption and decryption of the data. In order to make this algorithm much more secured, several symmetric cryptographic algorithms have been approved by NIST for the protection of sensitive data. The symmetric cryptographic algorithm combined with the use of hash function has been specified and approved in SP-800.

\subsection{Asymmetric Cryptographic Algorithms}
Asymmetric cryptographic algorithm (or public-key algorithm) uses a pair of keys, a public-key and a private-key which are related to each other. The public-key may be made public without reducing the security of the process, but the private-key must remain secrete if the cryptographic protection is to remain effective. 

\subsubsection{Digital Signatures Algorithms}
Digital signatures algorithms are used to provide identity authorization, integrity authorization, source authorization, and support for non-repudiation. Digital signatures are used in conjunction with hash functions The FIPS-186 specifies algorithms that are approved for the computation of digital signatures. It specifies several types of algorithms; RSA algorithm, the Elliptic Curve digital signature algorithm (ECDSA), and the Edwards Curve digital signature algorithm (EdDSA). The FIPS-186 also specifies additional requirements in each algorithms, which include the key sizes for each algorithms, and methods for generating the key pairs.

\begin{enumerate}
\item 
{\textbf{RSA scheme}}: \\
The RSA scheme is the most simple public-key algorithm, which is approved in FIPS-186 for the generation and verification of digital signatures, and specified in PKCS-1 and RFC-9017. Since the RSA digital signature scheme has a mathematically hard problem, that is so called, the Integer Factorization Problem (IFP), most of the cryptocurrencies do not use the RSA digital signature scheme.

\item
{\textbf{ECDSA scheme}}: \\
The ECDSA digital signature scheme is approved and specified in FIPS-186, and is used to generate and verify the digital signatures based on the use of elliptic curves. Since the ECDSA digital signature scheme is more secured than the RSA scheme. most of the major cryptocurrencies, such as Bitcoin (BTC), Ethereum (ETH), Ripple (XRP), have adopted the ECDSA scheme.

\item
{\textbf{EdDSA scheme}}: \\
The EdDSA digital signature scheme is adopted in FIPS-186 and described in RFC-8032. While the ECDSA scheme requires the use of a random (unique) value for the generation of each signature, the EdDSA scheme is deterministic, that is, the unique value for the generation of each signature is computed by the private key. While it is not as widely adopted as ECDSA, some cryptocurrencies, such as Monero (XMR), Zcash (ZEC), Stellar (XLM), have adopted EdDSA for their digital signature algorithm. 
\end{enumerate}

\subsubsection{Key Establishment Schemes}
Asymmetric cryptographic algorithms are used to set up keys to be used in communicating between entities. The scheme is a set of transformations that provide a cryptographic service. The scheme is used in a protocol that actually performs the communication needed for the key establishment process. There are two types of approved key establishment schemes; discrete-log-based scheme and integer factorization scheme. SP-800 specifies key establishment schemes that use RSA algorithm is also approved to use for key establishment, as well as for the generation and verification of digital signatures. 

\begin{enumerate}
\item \textbf{Diffie-Hellman (DH) and MQV}: \\
SP-800 specifies key establishment schemes that use discrete-log-based algorithms. Two algorithms are approved for key agreement; Diffie-Hellman (DH) and MQV. 

\item \textbf{RSA}: \\
RSA algorithm can be used for key establishment, as well as for the generation and verification of digital signatures. Its use for key establishment is specified in SP-800, it specifies approved methods for both key agreement and key transport. 
\end{enumerate}


\newpage
\begin{thebibliography}{99}
\bibitem[S. Joshi, A. Choudhury, R.I. Minu (2023)]{joshi2023}
S. Joshi, A. Choudhury, R.I. Minu,
{\it 'Quantum blockchain-enabled exchange protocol model for decentralized systems',}
Quant.Inf.Proc. 22 (11), 404 (2023).

\bibitem[A. Naik, E. Yeniaras, G. Hellstern, G. Prasad, S. Vishwakarma (2023)]{naik2023}
A. Naik, E. Yeniaras, G. Hellstern, G. Prasad, S. Vishwakarma,
{\it 'From Portfolio Optimization to Quantum Blockchain and Security: A Systematic Review of Quantum Computing in Finance',}
arXiv:2307.01155

\bibitem[D. Stebila, M. Mosca, N. Lutkenhaus (2010)]{Steb2010}
D. Stebila, M. Mosca, N. Lutkenhaus, 
{\it 'The Case for Quantum Key Distribution',}
Quantum Communication and Quantum Networking, In: Sergienko, A., Pascazio, S., Villoresi, P. (eds) Quantum Communication and Quantum Networking. QuantumComm 2009. Lecture Notes of the Institute for Computer Sciences, Social Informatics and Telecommunications Engineering, vol 36. Springer, Berlin, Heidelberg.

\end{thebibliography}
\end{document}